\documentclass[twocolumn,showpacs,amsmath,amssymb,superscriptaddress]{revtex4-1}
\usepackage{amsmath,amssymb,color,latexsym,natbib,graphicx}
\usepackage{dcolumn}
\usepackage{subfigure}
\usepackage{bm}
\usepackage{color}
\def\ADD#1{{\textcolor{black}{#1}}}

\newcommand{\be}{\begin{equation}}
\newcommand{\ee}{\end{equation}}
\def\ba{\begin{eqnarray}}
\def\ea{\end{eqnarray}}

\def\ie{{\it i.e.}\ }

\def \pmbmath{\mathpalette\pmbmathaux}
\def \pmbmathaux#1#2{
         \pmbtext{$#1#2$}}
\def \pmbtext#1{\leavevmode
     \setbox0\hbox{#1}
     \kern-0,2pt \copy0 \kern-\wd0
     \kern0,4pt \copy0 \kern-\wd0
     \kern-0,2pt \raise0,3pt \box0 }
\def\bol{{\pmbmath{\Omega}}}
\def\bell{{\pmbmath{\ell}}}

\def\kk{{\bf k}}

\def\kkp{{\bf k}_{\perp}}

\def\diso{d\ppn d\qpn d\ppa d\qpa}
\def\kpn{k_{\perp}}
\def\ppn{p_{\perp}}
\def\qpn{q_{\perp}}
\def\kpa{k_{\parallel}}
\def\ppa{p_{\parallel}}
\def\qpa{q_{\parallel}}

\def\nnn{{\tilde n}}
\def\mmm{{\tilde m}}

\begin{document}
\renewcommand{\thesubfigure}{}

\preprint{1}

\title{Theory for helical turbulence under fast rotation}
\author{S\'ebastien Galtier}
\affiliation{Laboratoire de Physique des Plasmas, Ecole Polytechnique, F-91128 Palaiseau Cedex, France}

\date{\today}

\begin{abstract}
Recent numerical simulations have shown the strong impact of helicity on \ADD{homogeneous} rotating hydrodynamic turbulence. 
The main effect can be summarized through the following law, $n+\tilde n = -4$, where $n$ and $\tilde n$ are respectively 
the power law indices of the one-dimensional energy and helicity spectra. We investigate this rotating turbulence problem 
in the small Rossby number limit by using the asymptotic weak turbulence theory derived previously. We show that the 
empirical law is an exact solution of the helicity equation where the power law indices correspond to perpendicular (to the 
rotation axis) wave number spectra. It is proposed that when the cascade towards small-scales tends to be dominated by the 
helicity flux the solution tends to $\tilde n = -2$, whereas it is $\tilde n = -3/2$ when the energy flux dominates. The latter solution 
is compatible with the so-called maximal helicity state previously observed numerically and derived theoretically in the weak 
turbulence regime when only the energy equation is used, whereas the former solution is constrained by a locality condition. 
\end{abstract}
\pacs{47.10.-i, 47.27.eb, 47.27.Gs, 47.27.Jv, 47.32.Ef}
\maketitle

\section{Introduction}

Geophysical flows like oceans, Earth's atmosphere, and gaseous planets are strongly influenced by the Coriolis force. 
\ADD{
The adimensional Rossby number, $R_o$, measures the relative strength of this force: it is the ratio between the magnitude of the advection 
term in the Navier-Stokes equations to the Coriolis term and reads $R_o \sim U/(L\Omega)$, where $U$ is a typical velocity scale, $L$ a 
typical length scale and $\Omega$ the rotation rate. For large-scale planetary flows $R_o \simeq 0.05$--$0.2$ \citep{shirley}.}

\ADD{
Many papers have been devoted to rotating turbulence ($R_o \le1$) but because of the very different nature of the investigations 
(theoretical, numerical and experimental) it is not possible in general to compare directly the results obtained. From a theoretical point of 
view it is convenient to use a spectral description in terms of continuous wave vectors with the unbounded homogeneity assumption in order 
to derive the governing equations for the energy, kinetic helicity and polarization spectra \citep{Cambon89} (see also \cite{chakraborty,galtier09} 
for investigations in the physical space or \citep{dubrulle} for simple heuristic models). Although such equations introduce transfer terms which 
remain to be evaluated consistently, it is already possible to show with a weakly nonlinear resonant waves analysis \citep{Cambon89,waleffe2} 
the anisotropic nature of that turbulence with a nonlinear transfer preferentially in the perpendicular (to $\bol$) direction. For moderate Rossby 
numbers the eddy damped quasi-normal Markovian model may be used as a closure \citep{Cambon97}, whereas in the small Rossby number 
limit the asymptotic weak turbulence theory can be derived rigorously \citep{galtier03a}. In the latter case, it was shown that the wave modes 
($\kpa>0$) are decoupled from the slow mode ($\kpa=0$) which is not accessible by the theory, and the positive energy flux spectra were 
also obtained as exact power law solutions. The weak turbulence regime was also investigated numerically and it was shown in particular 
that the two-dimensional manifolds is an integrable singularity and the energy cascade goes forward \citep{bellet}. Note that recently the 
problem of confinement has been addressed explicitly in the inertial wave turbulence theory using discrete wave numbers \cite{scott}. 
Three asymptotically distinct stages in the evolution of the turbulence are found with finally a regime dominated by resonant interactions. }

\ADD{
Pseudo-spectral codes are often used to investigate numerically homogeneous rotating turbulence \cite{bardina,mininni10a}. 
Several questions have been investigated like the origin of the anisotropy or of the inverse cascade observed when a forcing is applied at 
intermediate scale $k_f$. However, according to the question addressed the results may be affected by the discretization and by finite-box 
effects at too small Rossby number and too long elapsed time \cite{smith2,bourouiba08,bourouiba12}. This seems to be the case in 
particular for the question of the inverse cascade mediated by the decoupling of the slow mode. 
For example, it was found that the one-dimensional isotropic energy spectrum $E(k) \sim k^{-x}$ may follow two different power laws with 
$2 \le x \le 2.5$ at small-scale ($k>k_f$) and $x \simeq 3$ at large-scale ($k<k_f$) \citep{smith1}. But it was also shown that the scaling at 
large-scale was strongly influenced by the value of the aspect ratio between the parallel and the perpendicular (to $\bol$) resolution, a 
small aspect leading to a reduction of the number of available resonant triads, hence an alteration of the spectrum with the restoration 
of a $k^{-5/3}$ spectrum for small enough vertical resolution.}

\ADD{
Several experiments have been devoted to rotating turbulence with different types of apparatus 
\citep{hop,jacquin,baroud,morize,vanbokhoven,lamriben}. Contrary to the theory and the simulation, it is very challenging to reproduce 
experimentally the conditions of homogeneous turbulence. Nevertheless, one of the main results reported is that the rotation leads 
to a bi-dimensionalisation of an initial homogeneous isotropic turbulence with anisotropic spectra where energy is preferentially 
accumulated in the perpendicular (to $\bol$) wave numbers $k_\perp$. Energy spectra with $x \ge 2$ were experimentally observed 
\citep{baroud,morize,vanbokhoven} revealing a significant discrepancy with the isotropic Kolmogorov spectrum ($x=5/3$) for non-rotating 
fluids. Note that the wave number entering in the spectral measurements corresponds mainly to $k_\perp$. Recently, direct measurements 
of energy transfer have been made in the physical space (by using third-order structure function) with the detection of an increase of 
anisotropy at small scales \cite{lamriben} which is in agreement with some theoretical studies \citep{jacquin,galtier03a,galtier09,bellet}. 
}

The role of helicity -- which quantifies departures from mirror symmetry \cite{moffatt} -- on rotating fluids has not been analyzed seriously 
until recently. \ADD{One reason is that it is difficult to measure the helicity production from experimental studies. The other reason}
is probably linked to the weak effect of helicity on energy in non-rotating turbulence. Indeed, in this case one observes a joint constant flux 
cascade of energy and helicity with a $k^{-5/3}$ spectrum for both quantities \cite{chen1}. Recently, several numerical simulations have 
demonstrated the surprising strong impact of helicity on fast rotating hydrodynamic turbulence \cite{mininni10a,mininni10b,teitelbaum09} 
whose main properties can be summarized as follows. When the \ADD{(large-scale)} forcing applied to the system injects only negligible 
helicity, the dynamics is mainly governed by a direct energy cascade compatible with an energy spectrum $E(\kpn) \sim \kpn^{-5/2}$ 
which is precisely the weak turbulence prediction \citep{galtier03a}. However, when the helicity injection becomes so important that the 
dynamics is mainly governed by a direct helicity cascade, different scalings are found following the empirical law:
\be
n+\nnn = -4 \, ,
\ee 
where $n$ and $\tilde n$ are respectively the power law indices of the one-dimensional energy and helicity spectra. This law 
may be explained by a phenomenology where the inertial wave time $\tau_{\rm I}$ is introduced but without taking into account 
anisotropy, \ie by considering $\tau_{\rm I} \sim 1/ \Omega$. 
\ADD{Note that the helicity flux injection used in simulations may appear as artificial in regards with what happens in experiments where 
helicity is often produced at relatively small-scale near the Ekman layers and then diffused towards the core of the flow \citep{godeferd99}.}

In this paper, we investigate the rotating turbulence problem in the small Rossby number limit by using the asymptotic weak turbulence 
theory previously derived \citep{galtier03a}. We will show that the empirical law is actually the exact solution of the helicity 
equation where now the indices are associated with the bi-dimensional axisymmetric energy and helicity spectra:
\be
E(\kpn,\kpa) \sim \kpn^{n} \vert \kpa\vert ^{m} \, , \quad 
H(\kpn,\kpa) \sim \kpn^{\tilde n} \vert \kpa\vert ^{\tilde m} \, . \label{seeksol}
\ee
It is proposed that when the cascade towards small scales is dominated by the helicity flux the solution tends to $\tilde n = -2$, whereas 
it is $\tilde n = -3/2$ when the energy flux dominates. 
The latter solution is compatible with the so-called maximal helicity state previously observed numerically and derived theoretically 
in the weak turbulence regime when only the energy equation is used.

\section{Weak turbulence solutions}

Fast rotation introduces in the problem a natural small parameter $\epsilon$ proportional to the Rossby number, from which it is possible 
to expand asymptotically the Navier-Stokes equations. This is classically done in the framework of the weak turbulence theory 
\cite{ZLF,nazarenko11}. The preliminary work to such asymptotic developments is the derivation of the dynamical equation 
for the wave amplitudes from which we can obtain the resonance conditions. Several properties of weak turbulence may be predicted when 
we study the resonance conditions \cite{Galtier2001}. For inertial waves, the nature of the triad interactions has already been investigated 
\ADD{with a helicity decomposition} \cite{waleffe2,galtier03a} and the analyses show that the fluid bi-dimensionalises under the effects of 
rotation \ADD{with a concentration of energy towards (but without reaching) the slow mode without invoking an inverse cascade. Note that 
there is an important difference with the triadic interactions encounter in Navier-Stokes equations since here the analysis is restricted to 
resonant triads which eventually cannot lead to the same classification \cite{biferale}.}
From the wave amplitude equation we may derive the wave kinetic equations governing 
the long-time behavior of second order moments (in our case the energy and helicity spectra). The achievement of any weak turbulence 
theory is the derivation of such equations with their properties, like the exact power law solutions. Contrary to a simple heuristic description, 
the weak turbulence theory offers the possibility to prove rigorously the validity of the power law spectra by checking the locality of the 
solutions. In addition, the sign of the fluxes may be found which gives the direction of the cascade. The latter point is particularly important, 
first, for the comparison with existing data and, second, because it is impossible to predict that from a phenomenology. 

The theory of weak inertial wave turbulence was derived in \cite{galtier03a}, it is therefore useless to re-derive it. 
We make the choice to directly recall the wave kinetic equations which describe the time evolution of the energy and helicity 
spectra in the anisotropic limit (which corresponds to the $\kpn \gg \kpa$ limit), namely: 
\begin{widetext}
$$
{\partial \over \partial t} {E_k \brace H_k} = {\Omega^2 \epsilon^2 \over 4} \sum_{s s_p s_q} \int
{s \kpa s_p \ppa \over \kpn^2 \ppn^2 \qpn^2} \left( {s_q \qpn - s_p \ppn \over \omega_k} \right)^2 
(s \kpn + s_p \ppn + s_q \qpn)^2 \sin \theta_q \, \delta(s\omega_k + s_p\omega_p + s_q\omega_q) 
$$
\be
{ E_q ( \ppn E_k - \kpn E_p ) + ( \ppn s H_k / \kpn - \kpn s_p H_p / \ppn ) s_q H_q / \qpn
\brace 
s \kpn \left[ E_q ( \ppn s H_k / \kpn - \kpn s_p H_p / \ppn) + ( \ppn E_k - \kpn E_p ) s_q H_q / \qpn \right] } \, 
\, \delta(\kpa + \ppa + \qpa) \, \diso \, .
\label{main}
\ee
\end{widetext}
In these equations $E_k \equiv E(\kpn,\kpa)$ and $H_k \equiv H(\kpn,\kpa)$ are respectively the axisymmetric bi-dimensional energy and 
helicity spectra ($\perp$ and $\parallel$ are the directions perpendicular and parallel to $\bol$), $\theta_q$ is the angle between the perpendicular 
wave vectors $\kkp$ and ${\bf \ppn}$ in the triangle made with ($\kkp$, ${\bf \ppn}$, ${\bf \qpn}$), 
\ADD{$\omega_k = 2 \Omega \kpa / k \simeq 2 \Omega \kpa / \kpn$}
is the inertial 
wave pulsation, and ($s$, $s_p$, $s_q$) are the directional polarities which are equal to $\pm$ (by definition $s \kpa \ge 0$). In Eq. (\ref{main}) the 
integration over perpendicular wave numbers is such that the triangular relation $\kkp + {\bf \ppn} +{\bf \qpn}={\bf 0}$ must be satisfied. 

The solutions of Eq. (\ref{main}) were previously derived for a turbulence dominated by a forward energy flux \cite{galtier03a}. In this case, only the 
energy equation is useful and the exact finite flux solutions -- obtained by applying a bi-homogeneous conformal transformation \cite{ZLF} -- are:
\ba
E(\kpn,\kpa) &\sim& \kpn^{-5/2} \vert  \kpa \vert ^{-1/2} \, , \label{KZK1} \\
H(\kpn,\kpa) &\sim& \kpn^{-3/2} \vert \kpa \vert^{-1/2} .
\label{KZK2}
\ea
Additionally, we can derive the statistically equilibrium solutions for which the energy flux is null. In this case, we have $n=1$, $m=0$, $\nnn=2$
and $\mmm=0$. (Note that this result was not given in \cite{galtier03a}.)

In a situation where the turbulence is dominated by a (forward) helicity flux it is necessary to consider the second equation for the helicity to 
derive the new exact power law solutions. We apply the bi-homogeneous conformal transformation (also called Kuztnesov-Zakharov transform) 
which consists in doing the following manipulation on the wave numbers $\ppn$, $\qpn$, $\ppa$ and $\qpa$:
\be
\begin{array}{lll}
\ppn &\to & \kpn^2 / \ppn \, , \\[.2cm]
\qpn &\to & \kpn \qpn / \ppn \, , \\[.2cm]
\vert \ppa \vert &\to & \kpa^2 / \vert \ppa\vert  \, , \\[.2cm]
\vert \qpa \vert &\to & \vert \kpa\vert \vert  \qpa\vert  / \vert \ppa \vert  \, . 
\end{array}
\label{trans}
\ee
We seek stationary solutions in the power law form (\ref{seeksol}) 
where the parallel components are taken positive. After substitution, transformation and simplification, we obtain finally:
\begin{widetext}
\be
{\partial \vert H_k \vert \over \partial t} \sim \int \, 
\left[E_0 \left(1 - \left({\ppn \over \kpn}\right)^{n-1} \left\vert{\ppa \over \kpa}\right\vert^{m} \right)
+ H_0 \left(1 - \left({\ppn \over \kpn}\right)^{\nnn-2} \left\vert{\ppa \over \kpa}\right\vert^{\mmm} \right) \right]
\label{KZt}
\ee
$$
\left( 1 - \left({\ppn \over \kpn}\right)^{n + \nnn + 4} \left\vert{\ppa \over \kpa}\right\vert^{m + \mmm + 1} \right) \diso \, , 
$$
\end{widetext}
where $E_0$ and $H_0$ are some complicated coefficients which are not relevant to write explicitly. 
The zero helicity flux solutions correspond to the cancellation of both members of the integral in the first line; it gives
$n=1$, $m=0$, $\nnn=2$ and $\mmm=0$. Note that these solutions are exactly the same as those derived 
from the energy equation. The most interesting solutions are, however, those for which the constant helicity flux is finite. 
In this case, we find the relations:
\ba
n+\nnn &=& -4 \, , \label{sol1} \\
m + \mmm &=& -1 \, . \label{sol2}
\ea
Since the cascade along the rotation axis is strongly reduced the most important scaling law is therefore the one 
for the perpendicular wave numbers. It is remarkable to see that the exact solution (\ref{sol1}) corresponds to the 
empirical law observed in many simulations where the helicity transfer dominates the energy transfer 
\cite{mininni10a,mininni10b,teitelbaum09}.

It is important to look at the domain of convergence of the integral to check the degree of locality of the power law 
solutions. After a lengthly calculation one arrives to the locality conditions:
\ba
-3 < n + m  < -2 \, , \label{d1} \\
-2 < \nnn + \mmm < -1 \, . \label{d2} 
\ea
We see that with the previous solutions (obtained from the energy or the helicity equations) we are at the border line
of the domain of convergence. However, we also know that this problem is strongly anisotropic and the inertial range 
in the parallel direction is strongly reduced with a cascade almost only in the perpendicular direction. Actually, if we neglect 
the inertial range in the parallel direction (which is equivalent to say $m=\mmm=0$) we obtain a classical result of weak 
turbulence in the sense that the power law indices of the exact solutions (\ref{KZK1})--(\ref{KZK2}) fall then at the middle 
of the domains of locality (\ref{d1})--(\ref{d2}). 
Finally, from the helicity equation it is possible to show that the helicity flux is positive which is compatible with a 
direct cascade.

\section{Discussion}

The main result of this work is the derivation of the exact relations (\ref{sol1})--(\ref{sol2}) for a constant and finite helicity flux. 
It is thought that the empirical law found numerically \cite{mininni10a,mininni10b,teitelbaum09} 
(and also explained with a simple isotropic phenomenology) \ADD{may be} the signature of the weak inertial wave turbulence 
regime. Additionally, the present study gives an interesting limit for the perpendicular scaling -- assuming that the parallel transfer 
is negligible. Indeed, when the helicity flux is negligible, we find the so-called maximal helicity state which is a particular 
solution of the Schwarz inequality $H(\kk) \le k E(\kk)$ \ADD{(note that in the weak turbulence limit the polarization term \cite{Cambon89} 
does not contribute)}. In this case we have $n=\nnn-1=-5/2$. As the helicity transfer increases the power law 
indices $n$ and $\nnn$ get closer. The condition of locality gives, however, a limit to this convergence, namely $n=\nnn=-2$. 
This limit is supported by a simple anisotropic phenomenology where the inertial wave time writes $\tau_{\rm I} \sim \kpn / (\Omega \kpa)$. 
The stochastic collisions of inertial waves leads to the following estimate for the helicity flux,
$\tilde \varepsilon \sim H / [\tau_{\rm eddy}^2/ \tau_{\rm I}]$, with $\tau_{\rm eddy} \sim 1/(\kpn v)$; hence, the helicity spectrum prediction,
$H(\kpn,\kpa) \sim \sqrt{\tilde \varepsilon \Omega} \kpn^{-2} \kpa^{-1/2}$. 
It is interesting to note that the large-scale atmospheric wavenumber spectra of wind \cite{nastrom}, with a power law index close to $-3$, 
are still compatible with the present theory of weak inertial wave turbulence when the parallel transfer is neglected. 

\ADD{
As a conclusion, it is interesting to open a discussion about the predictions for higher-order statistics. Assuming that weak inertial turbulence 
is monofractal (see \citep{nazarenko11} for a detailed discussion including intermittency), then we obtain the relation $\zeta^e_p = 3p/4$ in 
the non-helical case and possibly $\zeta^h_p = p/2$ in the helical case with by definition 
$\langle (\bf u (\bf x + \bf \bell) - \bf u (\bf x ))^p \rangle \sim \ell^{\zeta^{e,h}_p}$, where $\bf \bell$ has to be seen as a vector perpendicular to 
the rotation axis. Despite their own limitations exposed in the introduction, we note that both direct numerical simulations and experiments 
may exhibit the presence of self-similar scalings with exponents close to $\zeta^e_p$ \citep{vanbokhoven,mininni10b} or to $\zeta^h_p$ 
\citep{baroud,mininni09i}. Although the nature of this self-similar behavior has to be further investigated, it might be seen as an additional 
argument for a weak turbulence interpretation.}


\end{document}